\documentclass{article}

\usepackage{arxiv}

\usepackage[utf8]{inputenc} 
\usepackage[T1]{fontenc}    
\usepackage{hyperref}       
\usepackage{url}            
\usepackage{booktabs}       
\usepackage{amsfonts}       
\usepackage{nicefrac}       
\usepackage{microtype}      
\usepackage{lipsum}		
\usepackage{graphicx}
\usepackage{natbib}
\usepackage{doi}

\title{BiHeartS: Bilateral Heart Rate from multiple devices and body positions for Sleep measurement Dataset}

\author{ \href{https://orcid.org/0000-0002-2497-557X}{\includegraphics[scale=0.06]{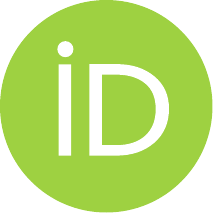}\hspace{1mm}Nouran Abdalazim}\\
	Faculty of Informatics\\
	Università della Svizzera italiana (USI)\\
	Via la Santa 1, Lugano 6900 \\
	\texttt{nouran.abdalazim@usi.ch} \\
	\And
	\href{https://orcid.org/0000-0003-3088-6132}{\includegraphics[scale=0.06]{orcid.pdf}\hspace{1mm}Leonardo Alchieri}\\
	Faculty of Informatics\\
	Università della Svizzera italiana (USI)\\
	Via la Santa 1, Lugano 6900 \\
	\texttt{leonardo.alchieri@usi.ch} \\
	\AND
   \href{https://orcid.org/0000-0002-5297-9893}{\includegraphics[scale=0.06]{orcid.pdf}\hspace{1mm}Lidia Alecci}\\
	Faculty of Informatics\\
	Università della Svizzera italiana (USI)\\
	Via la Santa 1, Lugano 6900 \\
	\texttt{lidia.alecci@usi.ch} \\
   \And
   \href{https://orcid.org/0000-0002-0882-2004}{\includegraphics[scale=0.06]{orcid.pdf}\hspace{1mm}Silvia Santini}\\
	Faculty of Informatics\\
	Università della Svizzera italiana (USI)\\
	Via la Santa 1, Lugano 6900 \\
	\texttt{silvia.santini@usi.ch} \\
}

\renewcommand{\headeright}{BiHeartS dataset}
\renewcommand{\undertitle}{BiHeartS dataset}
\renewcommand{\shorttitle}{Abdalazim et al.}

\hypersetup{
pdftitle={BiHeartS: Bilateral Heart Rate from multiple devices and body positions for Sleep measurement Dataset},
pdfauthor={Nouran Abdalazim, Leonardo Alchieri, Lidia Alecci, Silvia Santini},
pdfkeywords={BiHeartS dataset, Sleep quality, Wearable devices}}

\begin{document}
\maketitle

\begin{abstract}
	Sleep is the primary mean of recovery from accumulated fatigue and thus plays a crucial role in fostering people's mental and physical well-being. Sleep quality monitoring systems are often implemented using wearables that leverage their sensing capabilities to provide sleep behaviour insights and recommendations to users. Building models to estimate sleep quality from sensor data is a challenging task, due to the variability of both physiological data, perception of sleep quality, and the daily routine across users. This challenge gauges the need for a comprehensive dataset that includes information about the daily behaviour of users, physiological signals as well as the perceived sleep quality. In this paper, we try to narrow this gap by proposing \textbf{Bi}lateral \textbf{Heart} rate from multiple devices and body positions for \textbf{S}leep measurement (BiHeartS) dataset. The dataset is collected in the wild from 10 participants for 30 consecutive nights. Both research-grade and commercial wearable devices are included in the data collection campaign. Also, comprehensive self-reports are collected about the sleep quality and the daily routine.
\end{abstract}

\keywords{BiHeartS dataset \and Sleep Quality \and Wearable Devices}

\section{Introduction}


Sleep plays a crucial role in people's lives since it is the main source of recovery from daily fatigue \cite{gashi2022role}. 
Good sleep quality is thus instrumental to foster both physical and mental health \cite{alecci2022mismatch,lim2022assessing,khademi2019personalized,sano2018multimodal}.

Numerous studies in the literature, starting from the gold standard polysomnography (PSG), aim to assess individuals' sleep behavior. The development of wearable devices motivates their use in personal health monitoring systems \cite{roberts2020detecting}. 
Well-known commercial wearables, e.g., FitBit smartwatch and Oura smart ring, aim at providing insights about users' sleep behaviour \cite{gashi2022role}. They estimate different sleep parameters using physiological signals in real-world settings. Such estimates give the users insights about their sleep behaviour \cite{gashi2022role}. 


Nevertheless, the task of monitoring sleep behavior is far from simple. According to \citet{stone2020evaluations} and \citet{gashi2022role}, the use of physiological signals from wearable devices to monitor sleep quality indicators, e.g., total sleep time, latency, awakening and overall perceived sleep quality, is exceedingly difficult. 


This challenge originates from the intrinsic variability across users with respect to their daily behaviour, physiological signals and their perception about their sleep behaviour \cite{gashi2022role, khademi2018toward, alecci2022mismatch}. Specifically, sleep quality is hard to define since its determination is based on users' perception of their level of satisfaction about their sleep experience \cite{wang2021determinants}.  Such perception is influenced by human subjectivity and recall bias \cite{wang2021determinants}. 
This interpersonal variability hinders the sleep quality estimation task. To tackle this challenge, sleep quality estimation models must be trained on large number of users. This solution can help in handling the physiological signals variability. However, the variability in the perception of sleep quality still persists. 
Also, the variability in the daily behaviour, e.g., the physical activity during the day, the experienced stress level, can modify the perception of sleep quality.

The significant variability challenge serves as the catalyst behind the requirement for personalized sleep quality monitoring systems~\cite{gashi2022role}. This requirement highlights the importance of having an extensive dataset gathered in real-world settings in order to develop a sleep quality monitoring system suitable for practical, real-life situations.

To this goal, we propose a new sleep dataset entitled BiHeartS\footnote{Bilateral Heart Rate from multiple devices and body positions for Sleep measurement}. The data collection campaign includes 10 participants for 30 consecutive nights. We employ the research grade Empatica E4 wristband\footnote{\url{https://www.empatica.com/en-gb/research/e4/}} and the commercial Oura smart ring (Gen. 3)\footnote{\url{https://ouraring.com}}. Furthermore, we formulate the data collection protocol to acquire data suitable for various objectives, such as personalized sleep quality estimation, investigation of possible differences in physiological markers across body side, and comparisons between wearable devices.

\section{Related work}

Several researchers in literature have performed data collection campaigns to gather data about sleep behaviour through the use of wearable devices, each serving distinct objectives \cite{chinoy2022performance}, \cite{stone2020evaluations}, \cite{abdalazim2022}, \cite{sathyanarayana2016sleep}, \cite{rossi2020public} and \cite{gashi2022role}.

\citeauthor{chinoy2022performance} present an in the wild data collection campaign, aimed at comparing four commercial wearable devices. They included 21 participants for 7 nights. Similarly \citeauthor{stone2020evaluations} run an in the wild data collection campaign to compare nine commercial devices for 98 nights. The proposed dataset includes five participants. Both datasets are not publicly available. 
Moreover in \cite{abdalazim2022} we propose the HeartS dataset, which is collected from five participants in the wild with 3 wearable devices. The HeartS dataset is collected with the aim of comparing the physiological signals from three wearable devices and it is publicly available after signing a data sharing agreement.

The dataset proposed by \citeauthor{sathyanarayana2016sleep} contains actigraphy data from wearable medical devices, collected from 92 participants for 7 days. The aim of this data collection campaign is to study the feasibility of using the physical activity data during wake-up time as a predictor for sleep quality.

\citeauthor{rossi2020public} present the MMASH dataset, which contains 24 hour of continuous sensor signals namely: inter-beat intervals (IBI) signal, heart rate (HR) signal, and wrist accelerometer (ACC) signal. It also includes self-reports about daily activities as well as questionnaires about the chronotype, daily stress, sleep quality and anxiety. The MMASH dataset is publicly available for the scientific community. 

\citeauthor{gashi2022role} propose a dataset that is publicly available after signing a data sharing agreement. The dataset is collected in the wild for 30 days from 16 participants using Empatica E4 wristband along with self-reports about the sleep behaviour. The authors use the proposed dataset to explore the performance of personalized model in sleep/wake recognition and sleep quality score prediction tasks.

In light of the researchers' endeavors to provide the scientific community with comprehensive datasets about sleep behaviour using wearable devices, there is still a void in the availability of extensive multipurpose datasets that gather inclusive information about participants' daily behaviour, physiological patterns during sleep, and sleep quality perception. 

\section{BiHeartS Dataset}
\label{bihearts_dataset}
We collect a new sleep datatset entitled BiHeartS. We conduct a data collection campaign using both commercial and research-grade devices. The study is reviewed and approved by our Faculty's delegate for Ethics.

\subsection{Participants}
We recruit 10 participants (one female and nine males) of age from 20 to 30 years old.
Participants are asked to wear three wearable devices for 30 consecutive nights: (1)
Oura Ring (Gen. 3), 
which measures sleep with a Photoplethysmography (PPG) sensor; (2) Empatica E4 wristbands, one on each wrist, which is a research-grade wristband equipped with four sensors: Electrodermal Activity (EDA), Accelerometer (ACC), Skin Temperature (ST), and PPG sensors~\cite{siirtola2018using}.
The E4 wristband and the previous generation of Oura ring are adopted in several studies in the literature, e.g., \cite{chee2021multi, ghorbani2022multi, altini2021promise, stone2020evaluations, cakmak2020unbiased, assaf2018sleep, gashi2022role}. 

\subsection{Data Collection Procedure}
\label{sec:data_collection_procedure}
We formulate the data collection procedure to follow similar ones in the literature, e.g., \cite{gashi2022role, Sano2015_RecognizingAcademic, stone2020evaluations, abdalazim2022}. At the beginning, all participants sign an informed consent form. Then, we provide the wearable devices and the instructions needed to set up the applications for data synchronization. 
For collecting the ground truth, we rely on self-reports similar to \cite{gashi2022role}. We provide the participants with two options for logging their daily self-reports: pen-paper self-reports and RealLife Exp mobile application\footnote{\url{https://www.lifedatacorp.com}}. We split the self-reports to morning self-reports and evening self-reports. 

We advise the participants to put the pen-paper diary besides their beds, so they can easily remember to log their daily self-reports. Through the RealLife Exp mobile application, we send two notifications in the morning and two notifications in the evening to the participants as a reminder for the self-reports. 
To avoid recall bias, the participants cannot respond to the previous self-reports, i.e., the participants are allowed to log only self-reports of the current day.

Before the study, the participants answer a pre-study questionnaire, which integrates demographics questions and multiple validated questionnaires, i.e., the Pittsburgh Sleep Quality Index (PSQI) \cite{buysse1989pittsburgh}, the Morningness-Eveningness (Chronotype) questionnaire \cite{horne1976self}, the Big 5 inventory questionnaire for personality traits \cite{john1991big}, and the General Health questionnaire \cite{goldberg1979scaled}.

During the study, participants wear the Oura ring on their left hand along with the placement of two E4 wristbands, one on their left and one the right wrist every night. 
The participants wear the devices one hour before sleep and log the evening self-reports. 
The next day, participants complete the morning self-report then take off the devices one hour after waking up. During the day, participants synchronize the data from each device. To ensure integrity, we run daily checks for data synchronization compliance. After the study, the participants answer a post-study questionnaire that includes PSQI \cite{buysse1989pittsburgh}.

\subsection{Collected Data}

We collect two types of data, physiological signals using wearable devices and self-reports using RealLife Exp mobile application or pen-paper.

\paragraph{Physiological signals} From the Oura rings, we collect HR signal, Heart Rate Variability (HRV) signals, accessible through the Oura API\footnote{\url{https://cloud.ouraring.com/docs/sleep}}.
The Oura ring provides an average HR value for every five-minute interval throughout the sleep period. Similarly, for the HRV signal, the Oura ring offers an average HRV computed through the rMSSD method for each five-minute segment, while sleeping. 
Participants use the Oura mobile application\footnote{\url{https://support.ouraring.com/hc/en-us/articles/360058634153-How-to-Set-Up-the-Oura-App}} to synchronize the collected data to the Oura cloud dashboard. 
We use the Oura API to collect all of the participants' data. 
Beside the physiological signals provided by Oura ring, the API gives some parameters related to sleep behaviour, e.g., bedtime start, bedtime end, total bedtime, total sleep time, sleep time for each sleep phase, total awake time during sleep bedtime, latency and sleep efficiency.

From the Empatica E4 wristband, we collect five physiological signals, synchronously from both the left and the right wrist: HR, IBI, blood volume pulse (BVP), ST, EDA and ACC. Participants use the E4 manager desktop application\footnote{ \url{https://support.empatica.com/hc/en-us/articles/206373545-Download-and-install-the-E4-manager-on \newline -your-Windows-computer}} to synchronize the collected data to a pre-created study on the E4 connect website\footnote{\url{https://e4.empatica.com/connect/}}

\paragraph{Self-reports} We split the self-reports to morning self-reports and evening self-reports. The morning self-report includes questions about bedtime, latency, wake-up time, number of awakenings, a 10-points likert scale to report recovery score after sleep, and sleep quality. Participants also write any sleep disturbance reasons during the previous night. 

For the evening self-report, we include questions about the participant's behaviour during the day. The participants use a 10-points likert scale to report their sleepy score, stress score, and fatigue score. They also log the type and the duration of the conducted physical activity during the day. The evening self-report includes a 5-points likert scale for logging the daily physical health. 

The questions in both morning and evening self-reports align with the following validated questionnaires: PSQI questionnaire \cite{buysse1989pittsburgh}, General Health questionnaire \cite{goldberg1979scaled} and Physical Activity questionnaire~\cite{craig2003international}.

Our data collection campaign results in 148 sleep sessions, 234 morning self-reports and 213 evening self-reports. In total, for the personalized sleep quality estimation task, we have 146 sleep sessions along with the corresponding morning self-reports. 
Also, BiHeartS dataset has in total 91 complete sessions with Empatica E4 data collected from both sides of the body. While for the wearable devices comparison task, the dataset includes 135 complete sessions with data from the Oura ring and at least one Empatica E4 device. 

\section{Conclusion}
Sleep serves as the essential part of life for recuperating from accumulated fatigue. It has a vital function in fostering individuals' mental and physical wellness. 
The technological advancement of  wearable devices promotes their use in sleep quality monitoring systems. 
However, monitoring sleep quality is an arduous task given the inherent diversity among individuals in terms of their daily behaviour, physiological signals, and their self-perception of sleep quality. 
This challenge highlights the necessity for a comprehensive dataset that can be used to build a reliable sleep quality monitoring system, applicable in real world scenarios. In this paper, we propose the Bilateral Heart rate from multiple devices and body positions for Sleep measurement (BiHeartS) dataset. The dataset is collected in the wild from 10 participants for 30 consecutive nights. With BiHeartS, we collect extensive self-reports about participants' daily behaviour and sleep quality, as well as physiological signals from two wearable devices. BiHeartS is a multipurpose dataset that includes data that is feasible for various analysis tasks, e.g., personalized sleep quality estimation,  investigation of possible differences in other physiological markers across body side and comparison between wearable devices.

\section*{Acknowledgement}
This work has been supported by the Swiss National Science Foundation (SNSF) through the grant 205121\_197242 for the project ``PROSELF: Semi-automated Self-Tracking Systems to Improve Personal Productivity''.
\bibliographystyle{unsrtnat}
\bibliography{references}  

\end{document}